\documentclass[prb,showpacs,superscriptaddress,twocolumn,amssymb,amsfonts]{revtex4-1}
\usepackage{amsmath}
\usepackage{bm}
\usepackage{epsfig}
\usepackage{graphics}
\usepackage{dsfont}

\def\beq{\begin{equation}}
\def\eeq{\end{equation}}
\def\beqn{\begin{eqnarray}}
\def\eeqn{\end{eqnarray}}

\usepackage{color}

\begin{document}
\title{Condensation of Lattice Defects and Melting Transitions in Quantum Hall phases}
\author{Gil Young Cho}
\affiliation{Department of Physics, Institute for Condensed Matter Theory, University of Illinois, 1110 W. Green St., Urbana IL 61801-3080, U.S.A.}
\author{Onkar Parrikar}
\affiliation{Department of Physics, Institute for Condensed Matter Theory, University of Illinois, 1110 W. Green St., Urbana IL 61801-3080, U.S.A.}
\author{Yizhi You}
\affiliation{Department of Physics, Institute for Condensed Matter Theory, University of Illinois, 1110 W. Green St., Urbana IL 61801-3080, U.S.A.}
\author{Robert G. Leigh}
\affiliation{Department of Physics, Institute for Condensed Matter Theory, University of Illinois, 1110 W. Green St., Urbana IL 61801-3080, U.S.A.}
\author{Taylor L. Hughes}
\affiliation{Department of Physics, Institute for Condensed Matter Theory, University of Illinois, 1110 W. Green St., Urbana IL 61801-3080, U.S.A.}
\date{\today}

\begin{abstract}
Motivated by recent progress in understanding the interplay between lattice and electronic topological phases, we consider quantum-melting transitions of {\it weak} quantum liquid crystals, a crystal and a nematic phase, in which electrons form a quantum Hall state. In certain classes of Chern band insulators and quantum Hall phases, it has been previously demonstrated that there are topological Chern-Simons terms such as a Hall viscosity term and a gravitational Chern-Simons term for local lattice deformations. The Chern-Simons terms can induce anyonic statistics for the topological lattice defects and furthermore dress the defects with certain symmetry quantum numbers. On the other hand, the melting transitions of such liquid-crystalline orders are driven by the condensation of lattice defects. Based on these observations, we show how the topological terms can change the nature of the proximate disordered phases of the quantum liquid crystalline phases. We derive and study the effective dual field theories for the liquid crystalline phases with the geometric Chern-Simons terms, and carefully examine the symmetry quantum numbers and statistics of defects. We show that a crystal may go through a continuous phase transition into another crystal with the new discrete translational symmetries because the dislocation, the topological defect in the crystal, carries non-zero crystal momentum due to the Hall viscosity term. For the nematic phase, the disclination will condense at the phase transition to the isotropic phase, and we show that the isotropic phase may support a deconfined fractionally charged excitation due to the Wen-Zee term, and thus the isotropic phase and the nematic phase have different electromagnetic Hall responses. 
\end{abstract}


\maketitle

\section{Introduction} 

When a two-dimensional electron gas is subject to a strong magnetic field, it is well-known that the electronic energy spectrum consists of highly degenerate harmonic oscillator levels called Landau levels. In a partially-filled Landau level, the determination of the true ground state out of a massively degenerate set of levels is dominated by the repulsive interaction between the electrons since the kinetic energy is quenched. Because of the interactions, and the flat single-particle energy spectrum, one might naively expect for the electrons to form a crystalline state and break spatial translation symmetries. Surprisingly, the electrons often form an isotropic liquid-like state with a quantized electronic Hall conductance when certain conditions (e.g. particular fillings of the Landau levels) are met~\cite{Wen2004book, wenrev1, Wen95, Zhang1989, Lopez1991, Jain1989, Haldane1983}. For example, when the lowest Landau level is $\frac{1}{3}$-filled, the electrons form an incompressible fractional quantum Hall liquid described by the Laughlin wavefunction~\cite{Laughlin83}, and exhibit a fractionally quantized Hall conductance $\sigma_{xy} = \frac{e^{2}}{3 h}$.

In addition to the isotropic liquid phases there are also many types of phases in the quantum Hall regime which either exhibit spatial order or have a quantum Hall liquid coexisting with spatial order~\cite{Balents1996,Tevsanovic1989, Murthy2000, Mulligan2011, Maciejko2013, Kivelson1986PRL, Kivelson1987PRB, Fradkin1990, Fogler2002, Lee2002, Murthy2000, Radzihovsky2002}. The conventional spatial order can be classified by its broken spatial symmetries (e.g., in  2D there are two {\it continuous} translational symmetries and one continuous rotational symmetry that can be broken). A solid crystal phase is a phase with broken translational and rotational symmetries, e.g., a Wigner crystal phase. Crystal phases can melt to restore full translational symmetry through the condensation of  dislocations (N.B. we will use ``melting transition'' interchangeably with  phase transition)~\cite{kosterlitz1973,halperin1978,nelson1979,young1979}. The resulting intermediate phase with restored translation symmetry, but broken rotation symmetry, is called a nematic phase. A nematic phase can successively melt and restore its rotational symmetry by the condensation of disclinations. After this it becomes an isotropic liquid phase with full 2D spatial symmetry. 

We are interested in cases where a quantum Hall fluid remains incompressible but is subjected to coexisting spatial quantum order~\cite{Balents1996,Tevsanovic1989,Maciejko2013, Lee2002, Murthy2000}. For example,  spatial order on top of the incompressible quantum Hall fluid is expected~\cite{Balents1996, Lee2002} to emerge if the filling of the Landau level is slightly away from the `magic' filling and the dilute gas of  quasi-hole and quasi-electron excitations over the `magic' filling forms a liquid-crystalline or crystalline order. In principle,  a sufficiently large quadrupolar interaction between the electrons (which is the $L_z = \pm 2$ part of a rotationally symmetric interaction {\it e.g.,} a Coulomb interaction) may induce liquid-crystalline order~\cite{You_unpublished} while the Hall fluid remains incompressible. In other recent results it has been predicted that fractional Chern insulator states with higher Chern numbers can exhibit topological nematic phases\cite{barkeshli2012,wu2013} though we will not consider this system directly. Additionally, we note that we will not  address the perhaps more familiar examples of compressible quantum Hall smectic and nematic states\cite{Fradkin1990, Fogler2002,Radzihovsky2002}. 

More generally,  we focus on electrons in quantum Hall phases exposed to phase transitions in  `weak' quantum lattices~\cite{Kleinert2004, Zaanen2004, Zaanen2012, Beekman2013}. We use the term `weak' lattice to distinguish it from the conventional lattice formed by the ions in the solid which will not exhibit a quantum melting transition.  There are two possibilities to generate weak lattices in quantum Hall systems : (i) through spatial order parameters of the electrons or quasiparticles  exhibited in crystal or liquid crystalline phases or (ii) in {\it rotating} optical lattice systems~\cite{Demler2005, Demler2007, Lewenstein2005} where the potential wells can be adjusted to be weak and the structure can be tuned by hand. These weak lattices share the same symmetry with the classical ionic lattices and will have the same type of topological defects (dislocations and disclinations), but they have the advantage of the possibility of exhibiting melting transitions.  
The quantum lattice and its melting transition have also been shown to host other types of interesting physical phenomena such as  linearized Einstein-Cartan gravity~\cite{Zaanen2012, Kleinert2004}. We note that throughout this work we assume that the gapped electronic band structure (the Landau level cyclotron gap or many-body fractional quantum Hall gap) remains gapped and incompressible during the melting transitions of the background weak lattice. 


In such phases, it is natural to expect an interesting interplay between the geometric defects of the spatial order and the quantum Hall fluid. The connection between geometric degrees of freedom and the responses of quantum Hall phases has been intensely studied over the last five years based on older developments in the works of Wen and Zee\cite{Wen_Zee} and Avron {\it et. al.}~\cite{avron1995}. For our work the relevant results are the predictions that the quantum Hall states have Hall-like responses to geometric deformations~\cite{Wen_Zee,Read2009, Read2011, Hoyos2012, Bradlyn2012, Haldane2009, Hughes2011, Hughes2013, Gromov2014, avron1995}. In these works it was shown that the quantum Hall liquid responds to perturbations of the local geometry. The Wen-Zee term and gravitational Chern-Simons term encode the charge and angular momentum response of the quantum Hall state in the presence of curvature, including disclinations when rotation symmetry is broken by spatial ordering. On the other hand, the Hall viscosity indicates how the fluid will respond to strain-rates (velocity gradients) and, when coupled to a weak lattice, the response to dislocations.  


Similar to the pure charge response of a quantum Hall insulator, the geometric responses also have a Chern-Simons form. As such, the corresponding Chern-Simons terms in the hydrodynamic description will attach various quantum numbers to topological defects of the underlying spatial order, i.e., dislocations and disclinations. These quantum numbers can affect the statistics and the types of continuous phase transitions produced by condensation of these defects.
Thus, the quantum-melting transitions of a weak lattice controlled by the properties of the lattice defects are the main focus of this paper. We will separately consider a weak crystal and a nematic phase that coexists with a liquid-like quantum Hall state. We will show that the geometric Chern-Simons terms~\cite{Read2009, Read2011, Hoyos2012, Hughes2011, Hughes2013, Gromov2014, avron1995} can change the statistics and quantum numbers of the topological lattice defects  responsible for the melting transitions of the spatially ordered phases. We base our description of the relevant transitions on the dual theories of the relevant elastic theories~\cite{Zaanen2004, Beekman2013}. This description contains the low-energy physics of the spatially ordered phases and can characterize the nature of the defects that lead to proximate disordered phases.  We carefully examine when a continuous melting transition via condensation of the defects is available. When a continuous transition is available, we show that the topological order as well as the symmetries can change at a single transition. Furthermore, the topological defects become dressed with additional quantum numbers through the Chern-Simons terms. For example,  a dislocation in the crystal phase is dressed with crystal momentum~\cite{Hughes2011, Hughes2013}, and this implies that a crystal melts into another crystal with only discrete translational symmetry, rather than a nematic phase which has continuous translation symmetry.

The rest of the paper is organized as follows. In Section \ref{section:2}, we begin with the introduction to the duality transformation of a conventional bosonic theory coupled to a Chern-Simons term. In Section \ref{section:main1}, we 
consider the coexistence of a crystal phase and a quantum Hall state and develop a dual theory of the crystal. We will show that dislocations, i.e., the topological defect of the translation symmetry of the crystal, is bound to momentum in the presence of a quantum Hall state, similar to the response in the Chern insulator discussed in Refs. \onlinecite{Hughes2011,Hughes2013}. We carefully examine the quantum numbers of the dislocation and study the nature of the disordered phase of the crystal. In section \ref{section:main2}, we study the dual theory of a nematic phase coexisting with a quantum Hall phase and study the statistics and the electric charge of the disclination. We make some additional comments about possible melting transitions of the nematic phase and discuss some complications of the description of a complete theory of the nematic disordering transition that will be deferred to future work.

\section{Quantum Hall states with Lattice defects}\label{section:2}

As mentioned in the introduction, we will consider spatially ordered phases coexisting with quantum Hall electronic states. Our goal is to study the properties of the topological defects in a broken symmetry phase, including the types of phase transitions mediated by their condensation. Formally, the broken symmetries will be represented by their corresponding bosonic order parameters, and the order parameters will couple to the electronic degrees of freedom. After integrating out the gapped electronic degrees of freedom, one obtains an effective bosonic theory. For spatial symmetries the order parameters typically couple to the electrons as an effective spatial vielbein $ e^{a}_{i}$ or  metric $g_{ij}\equiv e^{a}_{i}e^{b}_{j}\delta_{ab}$ field.

In our cases of interest it has been shown that after integrating out the fermions one finds Chern-Simons type terms of the geometric fields that couple to the currents of the topological defects\cite{Read2009, Read2011, Hoyos2012, Hughes2011, Hughes2013, Gromov2014, avron1995}. The Chern-Simons terms could induce statistical transmutation of the lattice defects~\cite{Gromov2014} and/or dress the topological defects with certain quantum numbers~\cite{Hughes2011, Hughes2013}. Both of these effects play an important role in determining the available phase transitions. If the defects undergo statistical transmutation, then they can carry anyonic statistical angles. The anyonic statistics will have an important impact on the melting transitions of the spatial order parameter, {\it e.g.,} the order parameter cannot melt if the defects are not at the right ``filling'' in analogy with the anyonic quasi-particle condensation transitions in quantum Hall states. The melting of the spatial order itself occurs due to the proliferation/condensation of the appropriate topological defect. Thus, if the topological defects are dressed with additional quantum numbers (charge, spin, etc.), the defect condensate phase (i.e., the disordered phase) will break the symmetries associated with the attached quantum numbers. We will now discuss several interesting types of defect-mediated phase transitions, but we will first begin with reviewing the conventional theory of a bosonic order parameter coupled to a Chern-Simons term.


To understand the effects of the Chern-Simons terms, and to introduce the standard duality transformation on which we will  heavily rely in subsequent sections, we briefly review the theory of a conventional Abelian Chern-Simons term coupled to a boson field $\Psi$ which is charged under a non-compact U$(1)$ gauge field $A_{\mu}$~\cite{Wen2000, Lee1990, Wen1993, Chen1993,Pryadko1994fluctuation}: 
\begin{align}
L_{CS} &= \frac{1}{4\pi k} \varepsilon^{\mu\nu\lambda}  \alpha_{\mu} \partial_{\nu} \alpha_{\lambda} +  |(\partial_{\mu} - i\alpha_{\mu}-iA_{\mu}) \Psi|^{2} \nonumber\\
& +\frac{r}{2} |\Psi|^{2}+ V_{eff}(|\Psi|)+ \frac{1}{4{\tilde e}^{2}} (\varepsilon^{\mu\nu\lambda} \partial_{\nu}A_{\lambda})^2 + \cdots
\label{intro:CS}
\end{align}
From this action one can determine that the bosonic field $\Psi$ is bound to $k$ flux quanta of the gauge field $\alpha_{\mu}$ due to the presence of the Chern-Simons term ($\sim \varepsilon \alpha \partial \alpha$). Thus, $\Psi$  experiences statistical transmutation, i.e.,  if $k \in 2{\mathbb Z}$, $\Psi$ remains bosonic, otherwise, it can be fermionic or anyonic. The flux attachment, and its average field approximation, are expected to be reliable when the theory \eqref{intro:CS} describes the transition between the incompressible states of $\Psi$ such as a Mott insulator and a quantum Hall state. For now let us consider the case when the composite operator $\Psi$ (bound to $k$ fluxes of $\alpha_{\mu}$) is bosonic ($k \in 2{\mathbb Z}$). This allows us to study the transition between a Mott insulator and a fractional quantum Hall state at filling $\nu = \frac{1}{k}$. 

Then the theory in Eq. \ref{intro:CS} has two phases depending on the sign of $r$ (the mass of the boson field $\Psi$). When $r >0$, $\Psi$ is gapped and does not condense, thus it is the disordered phase of $\Psi$. In this case we can integrate out $\Psi$  and we find a gapless photon emerging from the non-compact U(1) gauge field $A_{\mu}$. Because the gauge field $A_{\mu}$ is non-compact, it is not confining and the photon is a stable gapless excitation. The effective action in this case is simply the Maxwell action: 
\beq
L^{(r>0)} = \frac{1}{4e^{2}} (\varepsilon^{\mu\nu\lambda} \partial_{\nu}A_{\lambda})^{2} + \cdots
\eeq 
This is the Mott insulator phase of the boson $\Psi$; the gauge field $\alpha_{\mu}$ is not in the ``flux-condensed'' phase~\cite{Zhang1989, Lopez1991, Wen1993, Chen1993, Pryadko1994fluctuation} and the Chern-Simons term for $\alpha_{\mu}$ in \eqref{intro:CS} can be safely ignored. This is clearly seen in the dual action of \eqref{intro:CS} which we will derive further below. 

On the other hand, if  $r<0$ then $\Psi$ will Bose condense and gain an expectation value. When $\Psi$ condenses, the action will develop a term $\sim |\Psi|^{2} \vert \alpha_{\mu} + A_{\mu}\vert^2$, i.e.,  there is a Higgs mass for the combination $\alpha_{\mu} + A_{\mu}.$  In the low-energy theory, this effectively sets $\alpha_{\mu} = -A_{\mu},$  and we find that the ordered phase of $\Psi$ is described by a single Chern-Simons theory: 
\beq
L^{(r<0)} = \frac{1}{4\pi k} \varepsilon^{\mu\nu\lambda} A_{\mu}\partial_{\nu}A_{\lambda} + \cdots
\label{intro:response}
\eeq
In this phase, there is no gapless photon from $A_{\mu}$ as there is  a topological mass gap from the Chern-Simons term.

Furthermore we can show that this phase $r<0$ supports fractional excitations. To exhibit this explicitly, we will use a dual theory. First we rewrite the boson field in terms of its phase and modulus $\Psi = \sqrt{\rho_0} \exp(i \phi)$. The phase field $\phi$ is compact over a range of $2\pi$, {\it i.e.,} $\phi \sim \phi + 2\pi$, which allows for topological defect configurations of vortices where
\beq
\oint d\vec{x} \cdot \nabla  \phi  = 2\pi {\mathbb Z}. 
\eeq
The kinetic term for the original bosonic field $\Psi$ from Eq. \ref{intro:CS} can be rewritten in terms of the modulus and the phase variables
\begin{align}
&\frac{1}{2} |(\partial_{\mu} - i\alpha_{\mu}-iA_{\mu})\Psi| = \frac{\rho_0}{2} |\partial_{\mu} \phi - \alpha_{\mu}-A_{\mu}|^{2} \nonumber\\ 
&\rightarrow  \frac{K^{2}_{\mu}}{2\rho_0} - K_{\mu}(\partial_{\mu}\phi - \alpha_{\mu}-A_{\mu})
\end{align}
where we have introduced the bosonic Hubbard-Stratonovich vector field $K_{\mu}$ in the above equation and assumed that $\rho_0$ is a constant (the fluctuation associated with the modulus is massive and can be safely integrated out in the low-energy theory). 

The phase variable $\phi$ can be separated into a smooth, single-valued part $\phi_{sm}$ and the singular part $\phi_{vor}$, i.e., $\phi = \phi_{sm} + \phi_{vor}$. The singular part of the phase variable encodes the vortex configurations of $\Psi$ via $\varepsilon^{ij}\partial_{i}\partial_{j} \phi_{vor} \neq 0$. The smooth part $\phi_{sm}$ is non-compact and can be integrated out to generate an equation of motion for $K_{\mu}$: 
\beq
K_{\mu} \partial_{\mu} \phi_{sm} \rightarrow \partial_{\mu}K_{\mu} =0, \quad K_{\mu} \equiv \frac{1}{2\pi} \varepsilon^{\mu\nu\lambda}\partial_{\nu}a_{\lambda}.
\eeq
By identifying $\frac{1}{2\pi} \varepsilon^{\mu\nu\lambda} \partial_{\nu}\partial_{\lambda} \phi_{vor} = J^{\mu}_{vor}$ as the {\it quantized} vortex current ($\int J^{0}d^2x \in {\mathbb Z}$), we end up with the dual theory of the theory in Eq. \ref{intro:CS}: 
\begin{align}
L_{CS} &= \frac{1}{4\pi k} \varepsilon^{\mu\nu\lambda}  \alpha_{\mu} \partial_{\nu} \alpha_{\lambda} - \frac{1}{2\pi} \varepsilon^{\mu\nu\lambda} a_{\mu}\partial_{\nu}(\alpha_{\lambda} + A_{\lambda})- a_{\mu}J^{\mu}_{vor} \nonumber\\
& + \frac{1}{4{\tilde g}^{2}} (\varepsilon^{\mu\nu\lambda} \partial_{\nu}a_{\lambda})^2+ \frac{1}{4{\tilde e}^{2}} (\varepsilon^{\mu\nu\lambda} \partial_{\nu}A_{\lambda})^2 + \cdots
\end{align}\noindent where ${\tilde{g}}^2=2\pi^2\rho_0.$
The theory is quadratic in  $\alpha_{\mu}$ so we can integrate it out and further simplify the low-energy theory in the long-wave length limit: 
\begin{eqnarray}
L_{CS}& = &\frac{k}{4\pi} \varepsilon^{\mu\nu\lambda} a_{\mu}\partial_{\nu} a_{\lambda} - a_{\mu}J^{\mu}_{vor} -\frac{1}{2\pi}\varepsilon^{\mu\nu\lambda} a_{\mu}\partial_{\nu}A_{\lambda}\nonumber\\ &+&  \frac{1}{4{\tilde e}^{2}} (\varepsilon^{\mu\nu\lambda} \partial_{\nu}A_{\lambda})^2 +\cdots
\end{eqnarray} 

To complete the duality transformation we need to elevate the vortex source current $J^{\mu}_{vor}$ to the level of a fluctuating  quantum field by introducing the soft-spin variable $\Phi$: 
\beq
J^{\mu}_{vor} \sim i \Phi^{*} (\partial_{\mu} \Phi) - i (\partial_{\mu} \Phi)^{*}\Phi.
\eeq 
Then we also need to supplement the theory with model kinetic and vortex-interaction terms to obtain:
\begin{align}
L_{CS} &= \frac{k}{4\pi} \varepsilon^{\mu\nu\lambda} a_{\mu}\partial_{\nu} a_{\lambda}  -\frac{1}{2\pi}\varepsilon^{\mu\nu\lambda} a_{\mu}\partial_{\nu}A_{\lambda} + \frac{1}{2} |(\partial_{\mu} - ia_{\mu})\Phi|^{2} \nonumber\\ 
&   +\frac{M^{2}}{2} |\Phi|^{2} + {\tilde V}_{eff}(|\Phi|) +  \frac{1}{4{\tilde e}^{2}} (\varepsilon^{\mu\nu\lambda} \partial_{\nu}A_{\lambda})^2 +\cdots
\label{intro:CS2}
\end{align}  which is the result we have been seeking. At this point, we can identify $\Phi$ as the vortex of the original boson $\Psi$ in \eqref{intro:CS} and it is anyonic with the statistical angle $\frac{\pi}{k}$ from \eqref{intro:CS2}. Thus the ordered phase $\Psi$ with the effective response \eqref{intro:response} is a fractionalized phase with the anyon $\Phi$. 

With Eq. \ref{intro:CS2} in hand we can try to understand the two different phases of the theory in Eq. \ref{intro:CS} from the dual action. The order-disorder transition of the original boson $\Psi$ from Eq. \ref{intro:CS} should occur when $\Phi$ condenses. In the ordered phase of the $\Phi$ field,  $\Phi$ condenses and $a_{\mu}$ develops a Higgs term and becomes massive. This results in an effective theory:  
\beq
L = \frac{1}{4{\tilde e}^{2}} (\varepsilon^{\mu\nu\lambda} \partial_{\nu}A_{\lambda})^{2} + \cdots
\eeq \noindent at low energy.
This is the same action that was found in the \emph{disordered} phase of $\Psi$. Thus we identify the ordered phase of $\Phi$ with the disordered phase of $\Psi$. Here, one can question how the field $\Phi$ can condense because the field $\Phi$ is anyonic with the Chern-Simons term of $a_{\mu}$. In the dual theory description of the transition, we have treated $\Phi$ as a bosonic field by ignoring the fluctuations of $a_{\mu}$ at the transition of $\Phi$. This was first used by Haldane~\cite{Haldane1983} and also used by Wen~\cite{Wen2004book} in the construction of the hierarchy fractional quantum Hall states. We note that one can also proceed with a $\Phi$ with an anyonic statistical angle to obtain the same hierarchy states. 

On the other hand, if $\Phi$ does not condense and is massive, then we can integrate out the $\Phi$ field to find  
\begin{align}
L &=  \frac{k}{4\pi} \varepsilon^{\mu\nu\lambda} a_{\mu}\partial_{\nu} a_{\lambda} -\frac{1}{2\pi}\varepsilon^{\mu\nu\lambda} a_{\mu}\partial_{\nu}A_{\lambda}  +\cdots \nonumber\\ 
& \rightarrow \frac{1}{4\pi k} \varepsilon^{\mu\nu\lambda} A_{\mu}\partial_{\nu}A_{\lambda} + \cdots 
\end{align} where we have also integrated out $a_\mu$ to obtain the second line. 
This is precisely the same effective theory \eqref{intro:response} appearing in the ordered phase of $\Psi$. Hence we identify the disordered phase of $\Phi$ as the ordered phase of $\Psi$. Thus, we have completed the dual description. 

Let us comment on the above dual description of the transition. First of all, exactly the same form of the theory~\cite{Wen1993} applies to the transition between quantum Hall states with different topological order. Second, the transition is continuous~\cite{Wen1993, Chen1993} in the large-N limit (where N is the number of flavors of the matter field coupled to the Chern-Simons gauge field) despite the presence of the Chern-Simons term, and hence is assumed to remain continuous in the small-N limit.  

Now that we have completed our review we will show that a similar field theory appears in the melting transition of a weak lattice with which a quantum Hall phase coexists. The melting transition is described by the condensation of the topological defects of the symmetry-broken phase. As in the transition between quantum Hall states, we will see that the geometric Chern-Simons terms restrict the operators that can condense at the transition. Furthermore, the defects are dressed with certain symmetry quantum numbers, just as the charge Chern-Simons term in \eqref{intro:CS} binds flux to a charge. Thus, after the condensation transition, the symmetries associated with the attached quantum numbers appear as the broken symmetries in the ordered phase of the defects.  Furthermore, because the melting transition is equivalent to the transition between fractional quantum Hall states, we will see that the topological order as well as the symmetry changes at a single transition. 

We will explain these ideas carefully as we discuss dislocation mediated melting of a weak lattice coexisting with a quantum Hall phase, and then move on to discuss the disordering of a nematic phase. 

\subsection{Case I. Dislocation Driven Transitions}\label{section:main1}
We will first consider the melting transition of a weak crystal lattice with no coexisting quantum Hall phase and later include the effects of a quantum Hall fluid that is coupled to the weak lattice. 
A crystal phase is a phase with broken translational and rotational symmetries. The low-energy theory is described by the local displacement vector $u_{a}({\vec x},t), a = 1,2$, and the quantized fluctuations of the displacement $u_{a}({\vec x},t)$ are the phonons of the crystal. The phonons are described by the Lagrangian   
\beq
L = \frac{1}{2g^{2}} \sum_{a=1,2} (\partial_{\mu} u_{a}({\vec x}))^{2}.
\label{ElasticTheory}
\eeq
There are two branches of phonons in two spatial dimensions  (one longitudinal, one transverse)  because the phonons are the Goldstone modes of the two broken continuous translational symmetries. A rotational Goldstone mode is absent from the low-energy physics, as it has been suggested~\cite{Beekman2013} that the rotational symmetry is {\it confined} in the crystal phase. Rotations of the spatial frame act to rotate the $u_a$ amongst themselves. Since there is no time-like component of $u_a$, the low energy theory will not have local Lorentz invariance, and there will be only two essentially Abelian currents involved in the low energy physics.

In the crystal there are two types of topological defects:  dislocations and disclinations~\cite{Zaanen2004, Hughes2013}. A dislocation is a {\it torsional} defect responsible for the restoration of the translational symmetry, whereas a disclination corresponds to the {\it curvature} defect responsible for the restoration of the rotational symmetry. The condensation of dislocations usually induces a transition toward a nematic phase in which the translational symmetries are restored. As the disclination is gapped both in the crystal and in the nematic phase, one can ignore the disclination~\cite{Zaanen2012} in the description of the initial transition between the crystal and the nematic phase. We will first study this transition.

To introduce a field theory for the transition, we identify the dislocation in terms of the displacement vector ${\vec u} = (u_{1}, u_{2})$ via 
\beq
\oint d{x}^{j}\partial_{j} {\vec u} = {\vec b}
\label{Dislocation}
\eeq
where ${\vec b} = (b_{1}, b_{2})$ is the Burgers' vector of the dislocations contained within the integration path. For convenience, we restrict ourselves to underlying spatial order which is a $C_2$ symmetric rectangular lattice generated by the two unit vectors $(l_{1}, 0)$ and $(0, l_{2})$ (we will further restrict to $C_4$ symmetry later when convenient). Then the Burgers' vector ${\vec b} =(b_{1}, b_{2})$ can be always represented as $n_{1} (l_{1},0) + n_{2}(0, l_{2})$ with $(n_{1}, n_{2}) \in {\mathbb Z}^{2}.$ The Burgers' vector represents two  quantized ``fluxes'' carried by the dislocation. The lattice thus sets a compactness relation for the displacement vector $u_{a} \sim u_{a}+ l_{a}$. Hence we split the displacement vector ${\vec u}({\vec x}, t)$ into the smooth part ${\vec u}_{sm} ({\vec x}, t)$ and the singular part ${\vec u}_{dis} ({\vec x}, t)$ which contains the information about the dislocation configuration via \eqref{Dislocation}. 

Now that we have identified the relevant topological defect we can dualize the theory \eqref{ElasticTheory}. The discussion here will parallel Ref. ~\onlinecite{Zaanen2004} and the previous subsection. We begin with the Hubbard-Stratonovich transformation of the elastic phonon Lagrangian
\begin{align}
L &= \frac{1}{2g^{2}} \sum_{a=1,2} (\partial_{\mu} u_{a}({\vec x}))^{2} \rightarrow  \sum_{a=1,2}\left(\frac{g^{2}}{2}(P^{a}_{\mu})^{2} - P^{a}_{\mu}\partial_{\mu}u_{a} \right)\nonumber\\
& \rightarrow \sum_{a=1,2} \left(\frac{g^{2}}{2 l^{2}_{a}} (\varepsilon^{\mu\nu\lambda}\partial_{\nu}{\cal K}^{a}_{\lambda})^{2} - \frac{1}{l_{a}}\varepsilon^{\mu\nu\lambda} {\cal K}^{a}_{\mu} \partial_{\nu} \partial_\lambda u^{a}_{dis}\right) \nonumber\\ 
& = \sum_{a=1,2} \left(\frac{g^{2}}{2 l^{2}_{a}} (\varepsilon^{\mu\nu\lambda}\partial_{\nu}{\cal K}^{a}_{\lambda})^{2} - {\cal K}^{a}_{\mu} j^{a}_{\mu}\right).
\end{align}
In the above equation, we have integrated out the smooth part ${\vec u}_{sm} \in {\mathbb R}^{2}$ to generate an equation of motion $\partial_{\mu} P^{a}_{\mu} = 0 \implies P^{a}_{\mu} \equiv \varepsilon_{\mu\nu\lambda} \partial_{\nu}{\cal K}^{a}_{\lambda}/l_{a}$ for $a=1,2$. We also have identified the quantized dislocation current ($\int d^2 xj^{a}_{0} \in {\mathbb Z}^2$):
\beq
j^{a}_{\mu} = \frac{1}{l_{a}} \varepsilon_{\mu}{}^{\nu\lambda} \partial_{\nu} \partial_{\lambda} u^{a}_{dis}, \quad a=1, 2.
\eeq  
We will now suppose that the dynamics of the system can be modeled by the possible condensation of  elementary  {\it dislocation fields}. For a particular dislocation, such a field would carry a `charge' corresponding to the  Burgers' vector of the dislocation. We suppose further that the dynamics is dominated by a pair of such fields corresponding to Burgers' vectors along the lattice basis directions. (Dislocations with other Burgers' vectors might be thought of as bound states of our elementary fields but we assume that these do not affect the vacuum structure in a significant way.) The dislocation currents $j^{a}_{\mu}$ can then be represented in terms of these quantum fields $\Phi_{a}$: 
\beq
j^{a}_{\mu} \sim i (\Phi_{a})^{*}\partial_{\mu} \Phi_{a} - i (\partial_{\mu} \Phi_{a})^{*}\Phi_{a}, \quad a= 1,2.
\eeq
The {\it dislocation fields}  $\Phi_a$ are the analogue of the vortex fields described earlier, and as implied by the form of the current, carry the `translational winding' associated with the dislocation. 

This completes the dual transformation of the elastic theory for the crystal \eqref{ElasticTheory} 
\begin{align}
L &= \sum_{a=1,2}  \left[ \frac{g^{2}}{2 l^{2}} (\varepsilon^{\mu\nu\lambda}\partial_{\nu}{\cal K}^{a}_{\lambda})^{2}  + \frac{1}{2} |(\partial_{\mu} - i{\cal K}^{a}_{\mu})\Phi_{a}|^{2} \right] \nonumber\\
& + \frac{r}{2} |\Phi_{1}|^{2} +  \frac{r}{2} |\Phi_{2}|^{2} + V_{eff}(|\Phi_{1}|, |\Phi_{2}|) + \cdots  
\label{Crystal:Dual}
\end{align}
In the dual theory \eqref{Crystal:Dual}, we have imposed $C_{4}$ rotation symmetry for convenience here, and thus the mass gaps for $\Phi_{1}$ and $\Phi_{2}$ are the same (furthermore, this fixes $l_{1} = l_{2} = l$). If one further relaxes this $C_{4}$ symmetry, then the mass terms for $\Phi_{1}$ and $\Phi_{2}$ do not need to be the same. With  $C_{4}$ symmetry, there are two phases depending on the sign of $r$ in \eqref{Crystal:Dual}.  

For $r>0$, the dislocations are massive and can be integrated out. In the low-energy and long wavelength limit, we find that there are two branches of the photons described by the two non-compact gauge fields ${\cal K}^{a}_{\mu},\; a= 1, 2$. This is the dual description of the two branches of the phonons in the original elastic theory \eqref{ElasticTheory}. Thus we identify the disordered phase of both $\Phi_{a},\; a= 1,2$ as the crystal phase. In the crystal, the dislocations are minimally coupled to the gauge fields and thus are logarithmically confined.

For $r<0$, the dislocations $\Phi_{a}, a=1,2$ condense and Higgs the gauge fields ${\cal K}^{a}_{\mu}, a= 1,2$. Because the dislocation condensate will restore the continuous translational symmetry, the ordered phase of the dislocations is the phase with translational symmetry. However, the disclinations are still gapped in this phase and thus the rotational symmetry is still broken. We identify this phase as the nematic phase. 

Now let us consider the impact of the electronic quantum Hall state coexisting with the crystal. We still assume that disclinations are gapped,  and we are interested in the regime where the low-energy physics is described by the phonons and the dislocations. For the topological Chern insulator\cite{Haldane1988} it was shown that  the response of the Chern insulator state coupled to the deformation vector ${\vec u}({\vec x}, t)$ can be derived~\cite{Hughes2013}, and it is the Hall viscosity term representing the response to torsion: 
\begin{align}
L_{\zeta_{H}} & = \sum_{a} \frac{\zeta_{H}}{2} \varepsilon^{\mu\nu\lambda} e^{a}_{\mu}\partial_{\nu} e_{a,\lambda}  \nonumber\\
&= -\sum_{a} \frac{\zeta_{H}}{2} \varepsilon^{\mu\nu\lambda} \partial_{\mu}u^{a}\partial_{\nu}\partial_{\lambda} u^{a} \nonumber\\
&= -\sum_{a} \frac{\zeta_{H}}{2} \varepsilon^{\mu\nu\lambda} \partial_{\mu}u^{a}_{dis} \partial_{\nu}\partial_{\lambda} u^{a}_{dis}.\label{eq:torsionCS}
\end{align}
We have used the fact that the vielbein $e^{a}_{\mu}$ and the deformation vector $u^{a}$ are related to each other by $e^{a}_{\mu} = \delta^{a}_{\mu}+ \partial_{\mu}u^{a}$ in linearized elasticity theory~\cite{Hughes2013}.  This Hall viscosity term encapsulates the response of the insulator to dislocations. Namely, to each dislocation there is an associated localized momentum density that is attached to the defect. While this was only shown for the Chern insulator, we will present arguments here that it should be a generic feature of quantum Hall states coupled to dislocation defects. 

\begin{figure}
\includegraphics[width=1\columnwidth]{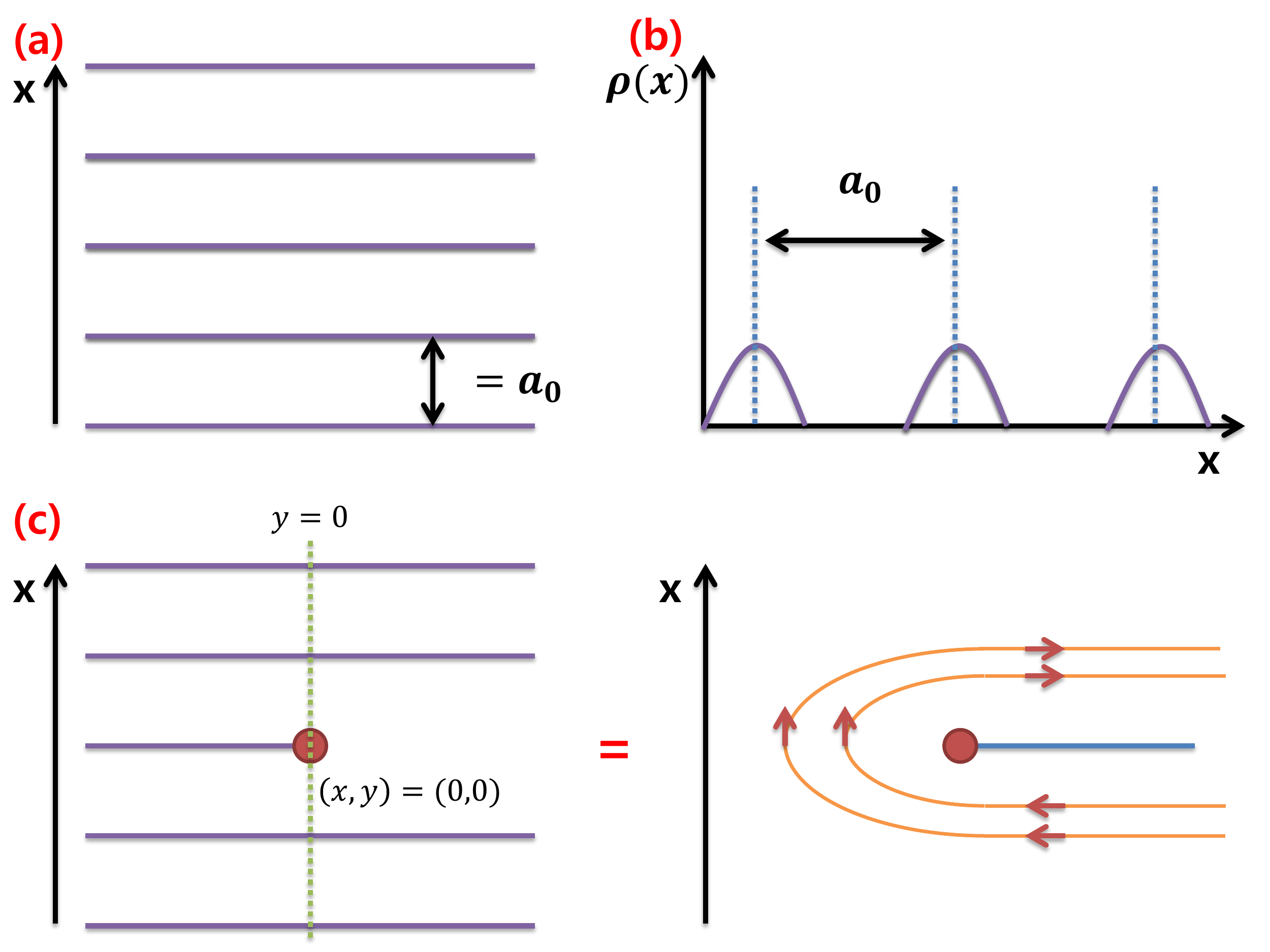}
\caption{Charge density wave over a quantum Hall fluid. (a) Charge density wave with periodicity $a_{0}$ along $x$ axis. (b) The charge density distribution along $\hat x$-axis. The straight lines represent the position of the extra charge due to the density wave. (c) A dislocation with the Burgers' vector $(a_{0}, 0)$ is introduced at the origin $(0,0)$ by removing charge. The charge distribution with the dislocation is ${\tilde \rho}(x,y) = \rho_{0} \sum_{n \in {\mathbb Z}} \delta (x- n a_{0}) - \rho_{0} \theta(-y)\delta(x)$. The position of the dislocation is identified with the red circle. From this charge distribution, it is straightforward to see that there is no net momentum current along $\hat y$-axis. The red arrows on the equipotential line represent the current which is proportional to the momentum density.}
\label{fig1}
\end{figure}

In a quantum Hall fluid, electrons in the bulk move chirally along equipotential lines of an electrostatic potential; this is a consequence of the Lorentz force.  The spatial order that coexists with the Hall fluid will only couple through such an electrostatic potential. When a dislocation is created in the crystal or charge density wave order over the quantum Hall fluid, an equipotential line will surround the dislocation (see Fig. \ref{fig1}c, for example) and this implies that there should be an electron current around the dislocation. For one-component Hall fluids the electric current is proportional to the local momentum density, and thus there is non-zero momentum density localized at the dislocation in the direction of the Burgers' vector. This is the same phenomenon predicted by Eq. \ref{eq:torsionCS}.

\begin{figure}
\includegraphics[width=1\columnwidth]{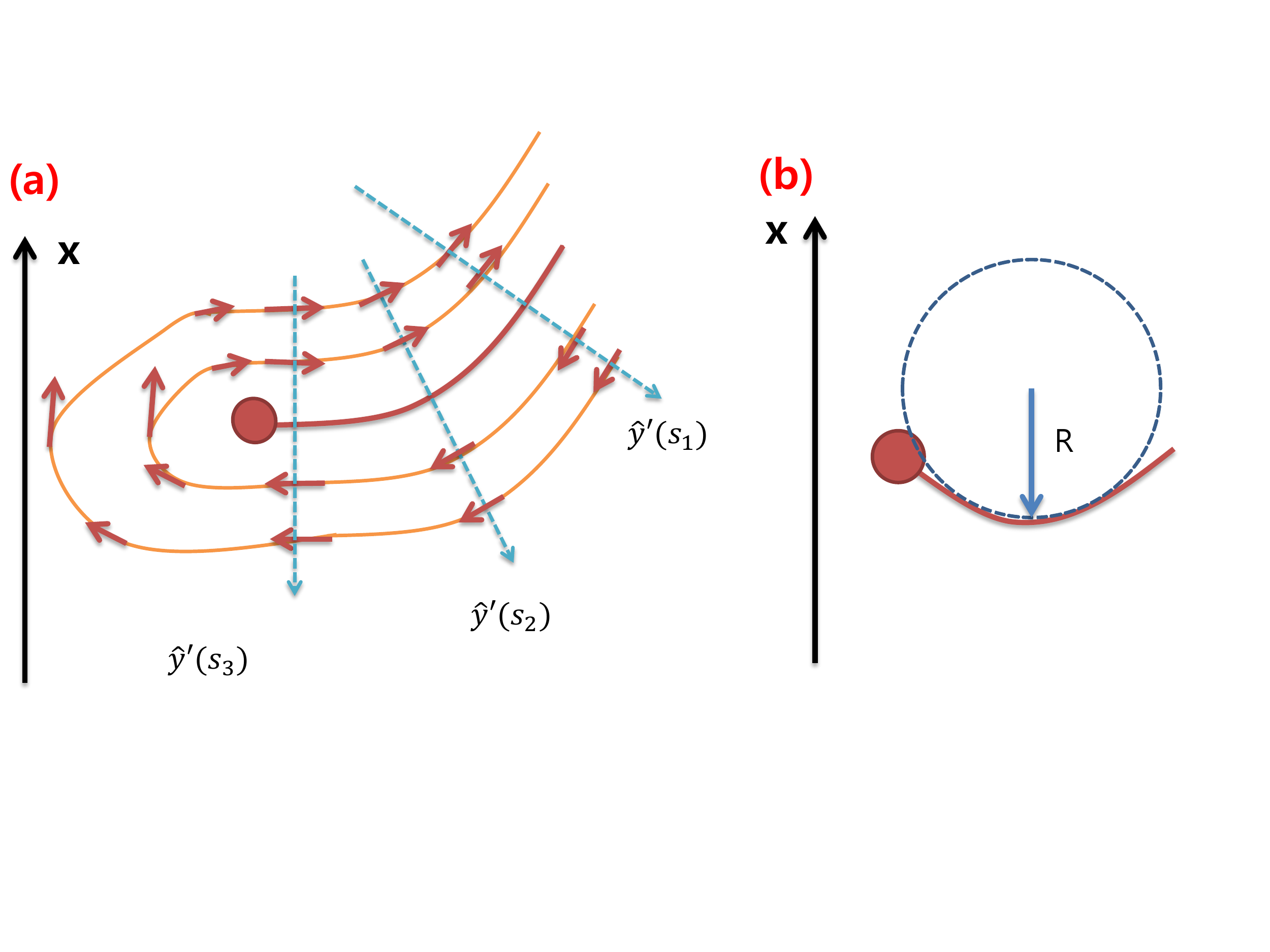}
\caption{Deformation of the Volterra cut for a dislocation.   (a) We introduce a local frame $(x', y')$ at each point where ${\hat x}' \propto \frac{d {\vec f}(s)}{ds}$ where ${\vec f} (s)$ parametrizes the location of the removed charge density. We introduce ${\hat y}'$ by requiring ${\hat y}' \cdot {\hat x}' =0$. (b) The curvature $\kappa(s)$ of ${\vec f}(s)$ at $s \in [0, 1]$ is $\frac{1}{R(s)}$ where $R(s)$ is the radius of the osculating circle at $s$.}
\label{fig2}
\end{figure}

 To be more explicit, let us consider a quantum Hall fluid coexisting with a  uni-directional charge density wave $\rho_{B}(x,y)$ with periodicity $a_{0} >> \ell_B$ (here $\ell_B$ is the magnetic length) is imposed as in Fig. \ref{fig1}a:
\beq
\rho_{B}(x,y) = \rho_{0}\sum_{n \in {\mathbb Z}} \delta (x- n a_{0}).
\eeq
We consider the case that the charge density wave generates a short-ranged electrostatic potential $V_{B}(x,y)$ for simplicity. The range of the potential we expect to be $l_{sc} \lesssim \ell_B$. With this assumption the background electrostatic potential $V_{B}(x,y)$ due to this charge distribution can be modeled as follows: 
\beq
(\partial^{2}_{x}+\partial^{2}_{y}-\frac{1}{l^{2}_{sc}}) V_{B}(x,y) = - \rho_{B}(x,y).
\eeq 
The background electric field due to the density wave is ${\vec E} = - {\vec \nabla} V_{B}$. We introduce a dislocation with the Burgers' vector $(a_{0},0)$ into the charge density wave. As demonstrated in Fig. \ref{fig1}c, the dislocation can be modeled  by removing charge along a half-line 
\beq 
\rho_{dis} (x,y) = -\rho_{0} \delta(x)\theta(-y). 
\eeq 
The total charge distribution with the dislocation over the Hall fluid is given by $\tilde{\rho}(x,y) = \rho_{B}(x,y) + \rho_{dis}(x,y)$. The electrostatic potential $V_{B}(x,y)$ will be modified accordingly as $V_{B} \rightarrow V_{B} + V_{dis}$ where $V_{dis}$ is the potential induced by a dislocation $\rho_{dis}(x,y)$. These quantities satisfy
\begin{align}
&(\partial^{2}_{x}+\partial^{2}_{y}-\frac{1}{l^{2}_{sc}}) (V_{B}+V_{dis}) = - \rho_{B}(x,y) - \rho_{dis}(x,y)  \nonumber\\
& \rightarrow (\partial^{2}_{x}+\partial^{2}_{y}-\frac{1}{l^{2}_{sc}}) V_{dis}(x,y) = - \rho_{dis}(x,y).
\end{align}
Because $\rho_{dis}(x,y) = \rho_{dis}(-x, y)$ then $V_{dis}(x,y) = V_{dis}(-x,y),$ and thus there is no net electric field along the $\hat x$-axis around the dislocation. This translates as no net electric current along the $\hat y$-axis due to the Lorentz force. On the other hand, there will be a non-zero net electric field along the $\hat y$-axis at the tip of the dislocation as illustrated in  Fig.\ref{fig1}c, and thus there is a localized momentum density at the dislocation in the direction of the Burgers' vector. The momentum density should be localized around the dislocation within a circle of radius $\sim l_{sc} \lesssim \ell_B$. 
 
In this argument we have used the fact that the ``string" attached to the point-like dislocation is a straight line. This is essentially a gauge choice, and we would like to see that this result is more general, and only depends on the point-like nature of the dislocation, not the attached string. So, with the result of this simple case in hand, we can consider a slightly more complicated case as in Fig. \ref{fig2} in which the string attached to the dislocation is curved, but with a curvature $\kappa << 1/l_{sc}$. Because the electrostatic potential should be screened on a length scale larger than $l_{sc}\sim \ell_B$, we can concentrate only on the region very close to the region of the removed charge (i.e., the string). This region is the area in which the current is localized and is much smaller than $1/\kappa$. 
For convenience, we parametrize the position of the removed charge (the string) by ${\vec f}(s), s \in [0,1]$. Then we can introduce a local frame ${\hat x'}$ and ${\hat y}'$ by ${\hat x}'(s) \propto \frac{d {\vec f}}{ds}$ (the tangent vector to the line ${\vec f}$) and ${\hat y}'(s) \perp {\hat x}'(s)$. Because we are interested in the region near the string within the length scale much smaller than the inverse of the curvature, we effectively have the same problem in the previous case (c.f. Fig. \ref{fig1}c) by ignoring any terms of $O(\ell_B\kappa)$:
\beq
(\partial^{2}_{x'}+\partial^{2}_{y'}-\frac{1}{l^{2}_{sc}}) V_{dis}(x',y') = - \rho_{dis}(x',y').
\eeq 
The important thing here is that the electron current is so tightly bound near the dislocation that it does not see the curvature, i.e., it cannot ``see" far enough away to distinguish the bending string from a straight line. Thus, $\rho_{dis}(x',y') = \rho_{dis}(x',-y')$, and we again have $V_{dis}(x',y') = V_{dis}(x', -y')$ and thus there is no net momentum along $\hat x'$-axis. On the other hand, it is apparent that there is a localized momentum density at the tip of the dislocation along the direction of the Burger's vector since at the tip, the local frame returns to just the initial $x,y$ axes. We note that effectively the assumption about the curvature is a restriction on the rate at which the curve bends so that when the string is thickened by including the region where the current is localized, then the thickened string will never intersect itself, i.e., different equipotential lines will not cross. It is possible this condition could be relaxed by microscopically resolving what occurs at the intersection between different equipotential lines, but we do not have an argument for that case.

From our arguments above, we claim that a dislocation coupled to a quantum Hall fluid traps momentum density. Hence, we will model the contribution of the quantum Hall fluid to the elastic action as 
\begin{equation}
L_{\zeta_{H}} = -\sum_{a}\frac{\zeta_{H}}{2} \varepsilon^{\mu\nu\lambda} \partial_{\mu}u^{a}_{dis} \partial_{\nu}\partial_{\lambda} u^{a}_{dis}.\label{eq:disCS}
\end{equation} There are two caveats about this expression. First, our arguments have shown that there should be a momentum bound to a dislocation but not what the magnitude of the momentum is in terms of the Burgers' vector. For Chern insulators, the proportionality between momentum and Burgers' vector is just the Hall viscosity, and we are assuming the same here. If it is not the Hall viscosity, it will generically be a response coefficient that is odd under time reversal symmetry, has the units of Hall viscosity, and is tunable by the magnetic field. These are the only properties on which our analysis relies so for simplicity we will take the coefficient to simply be the Hall viscosity. Second, we have not proven for the Hall fluid that this should take a relativistic form as it does in the Chern insulator, however introducing different coefficients for the various terms will not qualitatively affect our results so we will take the simplest case where they are all equal and we recover the relativistic form. We can proceed to invert this Hall viscosity term by introducing an auxiliary field ${\cal G}^{a}_{\mu}, a =1,2$ to write the Chern-Simons term coupled to the dislocation
\begin{align}
L_{\zeta_{H}} &= -\sum_{a}\frac{\zeta_{H}}{2} \varepsilon^{\mu\nu\lambda} \partial_{\mu}u^{a}_{dis} \partial_{\nu}\partial_{\lambda} u^{a}_{dis} \nonumber\\
& = -\sum_{a}\frac{\zeta_{H}l^{2}}{2}\varepsilon^{\mu\nu\lambda} \partial_{\mu}\left(\frac{u^{a}_{dis}}{l}\right) \partial_{\nu}\partial_{\lambda} \left(\frac{u^{a}_{dis}}{l}\right) \nonumber\\
& = \sum_{a} \left[ \frac{1}{2\zeta_{H}l^{2}}\varepsilon^{\mu\nu\lambda}{\cal G}^{a}_{\mu}\partial_{\nu}{\cal G}^{a}_{\lambda} - {\cal G}^{a}_{\mu} \varepsilon^{\mu\nu\lambda} \left( \frac{\partial_{\nu}\partial_{\lambda} u^{a}_{dis}}{l} \right) \right] \nonumber\\
& = \sum_{a} \left[ \frac{1}{2\zeta_{H}l^{2}}\varepsilon^{\mu\nu\lambda}{\cal G}^{a}_{\mu}\partial_{\nu}{\cal G}^{a}_{\lambda} - {\cal G}^{a}_{\mu}j^{a}_{\mu} \right]
\label{ViscosityChernSimonsTheory}
\end{align}\noindent where we have re-scaled by the lattice constant $l$ of the underlying weak square lattice. 
It is apparent from this result that the statistics of  a dislocation depends on the Hall viscosity $\zeta_{H}$. If we drag a dislocation around another dislocation it is translated by the Burgers' vector $b^a.$ It thus picks up a relative phase $\exp(ip_a b^a)$ where $p_a$ is the momentum carried by the dislocation traveling around the loop. Since the amount of momentum carried by the dislocation is proportional to $\zeta_H l$ and the length of the Burgers' vector is an integer in units of $l$ then the statistical phase depends on $\zeta_H l^2.$ In quantum Hall fluids, or a Chern insulator, the Hall viscosity $\zeta_{H}$ can be computed in  linear response theory~\cite{Read2009, Read2011, Hoyos2012, Hughes2011, Hughes2013, Gromov2014, avron1995}. 
\beq
\zeta_{H} = \frac{1}{8\pi \ell_{B}^{2}}
\eeq 
for some length scale $\ell_{B}$ set by the electronic structure in the Hall fluid/Chern insulator (note we have set $\hbar=1.$) In the quantum Hall fluid $\ell_B$ is the magnetic length, and in a Chern insulator it is a scale set by the bulk insulating gap. We will introduce a dimensionless number $k_{eff}= \frac{\zeta_{H}l^{2}}{2\pi}$ to rewrite Eq. \eqref{ViscosityChernSimonsTheory} in a more conventional form:
\beq
L_{\zeta_{H}} = \sum_{a} \left[ \frac{1}{4\pi k_{eff}} \varepsilon^{\mu\nu\lambda}{\cal G}^{a}_{\mu}\partial_{\nu}{\cal G}^{a}_{\lambda} - {\cal G}^{a}_{\mu}j^{a}_{\mu} \right].
\eeq

Now we can introduce the soft-spin field $\Phi_{a}$ to describe the dislocation, and we end up with the dual theory coupled with the Chern-Simons theory emergent from the Hall viscosity term. 
\begin{align}
L &= \sum_{a=1,2}  \left[ \frac{g^{2}}{2 l^{2}} (\varepsilon^{\mu\nu\lambda}\partial_{\nu}{\cal K}^{a}_{\lambda})^{2}  + \frac{1}{2} |(\partial_{\mu} - i{\cal K}^{a}_{\mu}-i{\cal G}^{a}_{\mu})\Phi_{a}|^{2} \right] \nonumber\\
& + \sum_{a} \left[ \frac{1}{4\pi k_{eff}} \varepsilon^{\mu\nu\lambda}{\cal G}^{a}_{\mu}\partial_{\nu}{\cal G}^{a}_{\lambda} \right]  + \frac{r}{2} |\Phi_{1}|^{2} +  \frac{r}{2} |\Phi_{2}|^{2} +\cdots
\label{Crystal:DualCS}
\end{align}
This is formally equivalent to the theory that we have studied in \eqref{intro:CS}. However there is a subtle difference in that $k_{eff}$ is not necessarily quantized. One might worry that non-quantized $k_{eff}$ implies that the gauge theory has irrational topological degeneracy $k_{eff}$ on $T^{2}$. However we see that the dislocation $\Phi_{a}$ is logarithmically confined due to the gapless photon ${\cal K}^{a}_{\mu}$, and thus there is no topological degeneracy associated with $k_{eff}$ in the crystal phase. Equivalently, the Wilson loop operators that transform from one ground state to a degenerate ground state cost energy and will split the ground-state degeneracy.

We note that the dislocation field $\Phi_{a}$ is bosonic if the lattice constant $l$ and the magnetic length $\ell_{B}$ have the correct ratio. In the context of the quantum Hall effect \eqref{intro:CS}, this is equivalent to saying that we should have the correct filling factor $k \in 2{\mathbb Z}$ for the electrons to have a transition to a stable fractional quantum Hall state at filling $\nu= \frac{1}{k}$. When we are at the correct {\it filling factor} $k_{eff}$ for the dislocations, we may have a transition from the crystal phase to another phase via the condensation of the dislocations. The condition to have a bosonic excitation can be derived: 
\beq
k_{eff} \in 2{\mathbb Z}  \leftrightarrow \ell_B = \frac{l}{ 4\pi\sqrt{2{\cal{N}}}}, \quad {\cal{N}} \in {\mathbb Z}.\label{eq:disfilling}
\eeq

We have noted that the Hall viscosity term \eqref{eq:disCS} binds crystal momentum to a dislocation. This is analogous to the fact that the Chern-Simons term in \eqref{intro:CS} binds electric charge to the flux of the gauge field. It can be shown that the dislocation of the ``flux'' $(n_{x}, n_{y}) \in {\mathbb Z}^{2}$ carries the momentum ${\vec p}$~\cite{Hughes2011, Hughes2013} (``charge'' bound to  ``flux''):
\beq
{\vec p} = \frac{2\pi k_{eff}}{l} \left(n_{x}, n_{y}\right).
\label{momentum}
\eeq
Because the momentum is carried by the dislocation, a condensate of dislocations will break the continuous translational symmetry with a period (the size of the unit cell) set by the momentum \eqref{momentum}. This should be sharply distinguished from the conventional melting transition of the crystal in which the dislocation condensation fully restores the translational symmetry. We should stress that the original discrete symmetry generated by the two unit vectors $(l,0)$ and $(0,l)$ is a subset of the new discrete translational symmetry set by \eqref{momentum}. Nevertheless, it is remarkable that we cannot restore the continuous translational symmetry by condensation of the dislocations.

Furthermore, we also notice that the dual theory \eqref{Crystal:DualCS} is equivalent to the theory \eqref{intro:CS} of the transition between a Mott insulator and a fractional quantum Hall state. Let us consider the case of $k_{eff} = 2{\cal{N}}, {\cal{N}} \in {\mathbb Z}$ to describe a continuous transition from a disordered phase of the dislocations to the condensed phase of the dislocations. If we assume that dislocations in  both directions condense simultaneously then the condensate in \eqref{Crystal:DualCS} produces a Higgs mass for the combination ${\cal G}^{a}_{\mu} + {\cal K}^{a}_{\mu}$ and sets ${\cal G}^{a}_{\mu} \approx -{\cal K}^{a}_{\mu}$ (for $a=1,2$) in the low energy regime. Then we see that $A^{a}_{\mu}$ receives a Chern-Simons term which was originally in terms of ${\cal K}^{a}_{\mu}$:
\beq
L  = \frac{1}{4\pi k_{eff}} \varepsilon^{\mu\nu\lambda}{\cal G}^{a}_{\mu}\partial_{\nu} {\cal G}^{a}_{\lambda} + \cdots 
\label{CS:Disordered}
\eeq 
The Chern-Simons term then generates the mass gap for the photon emergent from the gauge field ${\cal G}^{a}_{\mu}$. Additionally, it is straightforward to see that the ordered phase of the dislocations supports a {\it deconfined} fractional excitation which is the vortex of $\Phi_{a}$. Because the dislocation $\Phi_{a}$ is the order parameter for the new crystalline order with the lattice period dictated by \eqref{momentum}, we expect that the vortex of $\Phi_{a}$ should be again the dislocation in the new crystal phase. This phase corresponds to the ``torsion-condensed'' phase,  which is analoguous to the ``flux-condensed'' phase,{\it i.e.,} a fractional quantum Hall phase. In the fractional quantum Hall phase, the excitations are the (magnetic) flux, and thus the excitation (which is the vortex of $\Phi_a$) in this torsion-condensed phase is the source of the torsion, or equivalently a dislocation.

In summary, we find that the ordered phase of the dislocations is again a crystal phase which breaks continuous translational symmetry. Furthermore it has topological order with deconfined dislocations excitations. The disordered phase of the dislocation is trivially the original crystal phase.

\paragraph*{\ Explicit Example:} We now consider a specific example to illustrate the contents of the field theory we developed.  Let
\beq
l_{x} = l_{y} = 4\pi \ell_{B} \sqrt{2}.
\eeq  
From Eq. \ref{eq:disfilling}  this implies that we are at the correct ``filling'' (in analogy with the quantum Hall states) to have a bosonic dislocation in the excitation spectrum, i.e.,   
\beq
k_{eff}^{x} = k_{eff}^{y} = 2.
\eeq
Hence the minimal dislocations, with the ``fluxes'' $(n_{x}, n_{y}) = (1,0)$ and $(0,1),$ are bosonic. 

Now, consider the theory where the $(1,0)$ dislocation condenses. This dislocation carries a crystal momentum 
\beq
{\vec p} = \left(\frac{2\pi k^{x}_{eff}}{l_{x}}, 0\right) = \left(\frac{2\pi \times 2}{4 \pi \ell_B \sqrt{2}}, 0\right) = \left(\frac{1}{\ell_B\sqrt{2}},0\right). 
\eeq
The momentum sets a real space vector ${\vec R}$ such that ${\vec p}\cdot {\vec R} = 2\pi$, which defines a new unit cell in the ordered phase of $\Phi_{x}$. 
\beq
{\vec R} =\left(\frac{2\pi}{1/\ell_B\sqrt{2}}, 0\right) = \left(2\pi \ell_B \sqrt{2}, 0\right) =\left( \frac{l_{x}}{2}, 0\right). 
\eeq
Notice that we would usually expect for the condensation of the dislocation $(1,0)$ to restore the translational symmetry {\it completely} along the $x$-direction in the absence of the Hall viscosity term. However, we see that we cannot restore the continuous translational symmetry by condensation of the dislocation but end up with another crystal with half the period. 

\subsection{Case II. Disclination Driven Transitions}\label{section:main2}
Now we consider a background nematic phase with continuous translation symmetry and broken rotational symmetry. Let us review the relevant properties of the nematic phase and disclinations. The  continuous rotation symmetry of two-dimensional Euclidean space is broken down to a discrete subgroup $C_{n},$ or possibly completely broken. We will focus on the case with a discrete rotational symmetry $C_{n}, n \in 2, 3, 4, 6.$ For example, a triangular lattice can melt into a nematic phase with $C_{6}$ rotational symmetry, namely a hexatic state\cite{kosterlitz1973,halperin1978,nelson1979,young1979}. One consequence of the broken {\it continuous} rotational symmetry is the presence of a single Goldstone mode. Another consequence is the existence of a topological defect, namely a disclination. Ultimately the proliferation of disclinations is responsible for the restoration of the continuous rotational symmetry to return to an isotropic liquid state. In the presence of $C_n$ rotational symmetry a disclination carries a {\it quantized} Frank angle $\theta_{F} = \frac{2m\pi}{n}, m = 1, 2, \cdots n-1$. Generally the disclinations are coupled with the Goldstone mode and are logarithmically confined in the nematic phase. A  disclination with the Frank angle $\theta_F \in [0, 2\pi]$ can be modeled as a metric tensor for a point-source of curvature\cite{Vozmediano2008,Furtado2008} 
\beq
ds^{2} = dr^{2} + (1+\frac{\theta_F}{2\pi})^2r^{2}d\phi^{2}.
\eeq
The spin connection in this geometry has a non-zero component
\beq
\omega^{\phi}{}_{r} = \left(1+\frac{\theta_F}{2\pi}\right)d\phi.
\eeq
Thus, a disclination is a point source of curvature (as opposed to torsion) with an amount of curvature that is precisely $\theta_F.$ 

 While the nematic phase can support curvature defects, the torsion in a nematic phase will generically vanish. To understand the vanishing torsion in the nematic phase, we start by considering the crystal phase proximate to the nematic phase.  In the crystal phase, the torsion $T^{a}$ present in the system is proportional to the density of Burgers' vector $b^{a}$ (i.e., the dislocation density)~\cite{Zaanen2012, Kleinert2004, Beekman2013, Zaanen2004}. In a conventional crystal to nematic transition the dislocations are trivially bosonic and may condense to drive a transition toward the nematic phase. In fact, the nematic phase is the phase that the minimal dislocations with Burgers' vectors along the axes with discrete translation symmetry both condense.   Hence, the density of Burgers' vector in the nematic phase is zero, and thus the torsion vanishes. We can also conclude that torsion is still zero even in the presence of disclinations (which by definition carry a non-zero Frank angle, but can possibly carry a Burgers' vector),  because the torsion induced by the disclination can be screened by the dislocation condensate. This is true when the nematic is not coupled to an electronic system with a {\it weak} topological index of the electronic band structure~\cite{Teo2013,Teo2013b}.

To describe the nematic melting transition we introduce an angular variable $\theta \sim \theta + \frac{2\pi}{n} = \theta_{sm} + \theta_{vor}.$ In this splitting $\theta_{sm}$  describes the Goldstone mode and $\theta_{vor}$ describes the disclination in the nematic phase~\cite{Beekman2013} through the Lagrangian
\beq
L = \frac{1}{2g^{2}} (\partial_{\mu} \theta)^{2}.
\label{Nematic:ElasticTheory}
\eeq
Using the results from previous sections we use a similar method to dualize this elastic theory to obtain an effective theory for the disclination $\theta_{vor}$ coupled to a non-compact U(1) gauge field $a_{\mu}$: 
\begin{align}
L &= \frac{g^{2}}{8\pi^{2}n^{2}} (\varepsilon^{\mu\nu\lambda}\partial_{\nu}a_{\lambda})^{2} - \varepsilon^{\mu\nu\lambda}a_{\mu}\frac{n}{2\pi} \partial_{\nu}\partial_{\lambda} \theta_{vor}\nonumber\\
& = \frac{g^{2}}{8\pi^{2}n^{2}} (\varepsilon^{\mu\nu\lambda}\partial_{\nu}a_{\lambda})^{2} - a_{\mu}J_{\mu}, \quad J_{\mu} = \frac{n}{2\pi} \varepsilon^{\mu\nu\lambda}\partial_{\nu}\partial_{\lambda} \theta_{vor}.
\end{align} The non-compact gauge field $a_{\mu}$ is a dual description of the Goldstone mode. 
Here the disclination current with the minimal Frank angle $\theta_{F} = \frac{2\pi}{n}$ is quantized, $\int J_{0} \in {\mathbb Z}$. We introduce a soft-spin variable $\Phi$ to complete the dual transformation of the elastic theory \eqref{Nematic:ElasticTheory}.
\begin{align}
L &= \frac{g^{2}}{8\pi^{2}n^{2}} (\varepsilon^{\mu\nu\lambda}\partial_{\nu}a_{\lambda})^{2} + \frac{1}{2} |(\partial_{\mu} - ia_{\mu})\Phi|^{2} \nonumber\\ 
& + \frac{r}{2} |\Phi|^{2} + V_{eff}(|\Phi|) + \cdots 
\end{align}

Now, one can easily find the nature of the two phases separated by condensation of the disclination field $\Phi$. $r>0$ corresponds to the nematic phase in which the low-energy physics is described by a photon emerging from $a_{\mu}$. The photon is the dual description of the Goldstone mode of the broken rotational symmetry in the nematic phase. For $r<0$, the disclination condenses and the rotational symmetry is restored. This phase is the isotropic phase with the full continuous symmetry of two dimensional space. 

Armed with this knowledge, we now turn on a magnetic field to form the electronic integer quantum Hall state over the nematic phase of the quantum lattice. For concreteness, we consider the electrons in a uniform magnetic field in which only the lowest Landau level is filled.
We are interested in the linear response of electrons to the geometric perturbations which generate  curvature {\it without torsion}. The electronic degrees of freedom can be integrated out and we end up with the effective theory consisting of the electromagnetic Chern-Simons term, the Wen-Zee term~\cite{wenrev1, Wen_Zee}, and the gravitational Chern-Simons term~\cite{Hughes2013, Gromov2014}: 
\begin{align}
L &= \frac{1}{4\pi}\varepsilon^{\mu\nu\lambda}A^{em}_{\mu}\partial_{\nu}A^{em}_{\lambda} +   \frac{1}{4\pi}\varepsilon^{\mu\nu\lambda} A^{em}_{\mu}\partial_{\nu} \omega_{\lambda}\nonumber\\
&+\frac{1}{24\pi} \varepsilon^{\mu\nu\lambda} \omega_{\mu}\partial_{\nu} \omega_{\lambda}.  
\label{CS:Nematic}
\end{align}\noindent We note that usually in 2+1-d the spin connection is a matrix of one-forms $\omega_{\mu}^{ab}dx^{\mu}$ where $a,b= 0, 1, 2.$ For the non-relativistic quantum Hall effect the only important component is the rotational piece $\omega_{\mu}^{xy}.$ In Eq. \ref{CS:Nematic} $\omega_{\mu}\equiv \omega_{\mu}^{xy}.$ As we can see from this action, the electrons are sensitive to curvature, however, to proceed, we must assume that the electrons are sensitive to the sources of curvature supplied by disclinations in the nematic and that disclinations in the liquid crystal can be modeled as shown in the metric above. We expect that this assumption will only hold when the electrons are extremely strongly coupled to the nematic. It will almost certainly fail when the electrons are only weakly coupled because it is known that the electrons will not even see a quantized curvature from nematic disclinations\cite{Maciejko2013}. To be precise, the assumption we make is that curvature $\partial_\mu \omega_\nu-\partial_\nu\omega_\mu=(\partial_\mu\partial_\nu -\partial_\nu\partial_\mu)\theta_{vor}$ where $\theta_{vor}$ is the non-single-valued disclination field. 

We are now prepared to study the properties of the disclination-driven transition. First we rewrite the Wen-Zee term and the gravitational Chern-Simons term \eqref{CS:Nematic} in terms of the disclination field $\theta_{vor}$ 
\begin{align}
L &=  \frac{1}{4\pi}\varepsilon^{\mu\nu\lambda} A^{em}_{\mu}\partial_{\nu} \partial_{\lambda} \theta_{vor} +\frac{1}{24\pi} \varepsilon^{\mu\nu\lambda} \partial_{\mu}\theta_{vor}\partial_{\nu} \partial_{\lambda}\theta_{vor}  \nonumber\\ 
& = \frac{1}{2n} A^{em}_{\mu} J^{\mu} - \frac{6n^{2}}{4\pi} \varepsilon^{\mu\nu\lambda} \Omega_{\mu} \partial_{\nu} \Omega_{\lambda} - \frac{n}{2\pi} \varepsilon^{\mu\nu\lambda}\Omega_{\mu}\partial_{\nu}\partial_{\lambda}\theta_{vor}\nonumber\\
&= (\frac{1}{2n} A^{em}_{\mu}-\Omega_{\mu}) J^{\mu} - \frac{6n^{2}}{4\pi} \varepsilon^{\mu\nu\lambda} \Omega_{\mu} \partial_{\nu} \Omega_{\lambda}. 
\label{Nematic:effective}
\end{align}
We have inverted the gravitational Chern-Simons term by introducing an auxiliary field $\Omega_{\mu}$ and  used the quantized disclination current $J_{\mu} \in {\mathbb Z}$ with the smallest Frank angle $\frac{2\pi}{n}$. We can see that the disclination with the smallest Frank angle traps an electric charge $\frac{1}{2n}$ which is consistent with the result of  Ref.~\onlinecite{Gromov2014}. It is the identification of a quantized disclination current where we have had to use the strong-coupling assumption mentioned above.

From the theory \eqref{Nematic:effective}, we notice that the disclination is bound with a flux $\frac{2\pi}{6n^{2}}$ and has an anyonic statistical angle $\frac{\pi}{6n^{2}}$. To discuss the condensed phase of the disclinations, we first perform the flux-smearing approximation~\cite{Wen2004book, wenrev1, Wen95, Zhang1989, Lopez1991} and thus the disclinations will form  integer quantum Hall states with the filling $\nu_{eff} = 6n^{2} \in 2{\mathbb Z}$. This state can be thought as  $3n^{2}$ copies of the $\nu =2$ bosonic integer quantum Hall state~\cite{Senthil2013, Lu2012b, Lu2012a} which has been discussed in the context of the bosonic symmetry protected topological phases. The excitation in this state is a bosonic excitation~\cite{Senthil2013, Lu2012b, Lu2012a} with the electromagnetic charge $\frac{1}{2n}$ (this excitation {\it was} the disclination in the nematic phase and thus inherits the electric charge of the disclination). This implies that the electromagnetic Hall effect will be increased by 
\beq
\delta \sigma^{xy} = 6n^{2} \times (\frac{1}{2n})^{2} \frac{e^{2}}{h}= \frac{3e^{2}}{2 h}  
\eeq\noindent  in the ordered phase of the disclination.  
Furthermore, there will be no Goldstone mode associated with the broken rotational symmetry in the condensed phase of the disclination and thus the resulting phase will be an incompressible quantum Hall state. Here we have considered the case where only the lowest Landau level is fully filled, but it is straightforward to generalize our calculation to the other fillings. 

As there is experimental evidence for a nematic fractional quantum Hall state~\cite{Xia2011}, some of these predictions may carry over to that system, though the connection is not obvious. 
Away from the strong-coupling limit  we would expect the curvature induced by the disclination to be non-quantized, but it still determines the statistical angle and the charge of the disclination. A shot-noise experiment~\cite{Kim2005} or tunneling experiments might be applicable to detect the (non-quantized) statistics and charge of the disclination. This non-quantized statistics of the disclination could be clearly contrasted with the quasi-hole and quasi-electron excitations which alway carry quantized statistics in the quantum Hall states to distinguish the contributions.


Before finishing this section, we comment on a relevant issue for the case with electrons in a Chern Insulator studied in Refs. ~\onlinecite{Hughes2011,Hughes2013}. In these references, the authors found  geometric response terms including a gravitational Chern-Simons term in terms of the {\it non-abelian} spin connection $w^{ab}$~\cite{Hughes2013}: 
\beq
L = \frac{1}{96\pi}\varepsilon^{\mu\nu\lambda}\left[w_{\mu}{}^{a}{}_{b}\partial_{\nu}w_{\lambda}{}^{b}{}_{a} + \frac{2}{3} w_{\mu}{}^{a}{}_{b}w_{\nu}{}^{b}{}_{c}w_{\lambda}{}^{c}{}_{a} \right].\label{eq:gcsDirac}
\eeq\noindent Their calculations also show torsional response terms, but no clear candidate for a Wen-Zee term. 
For the curvature response, a disclination corresponds to a flux in $w^{xy}_{\mu}$ and traps a spin density $S_{z} = \psi^{\dagger} \sigma^{z}\psi$ of the electrons $\psi = (\psi_{\uparrow}, \psi_{\downarrow})^{T}$ in the Chern insulator.  With this in mind, it is not difficult to see that the condensation of disclinations will induce a spin/magnetic ordering in $S_{z}$ and hence the isotropic phase is a magnetized phase with non-zero $\langle S_{z} \rangle.$ It would be interesting to explore this in future work, as well as the unresolved issue of how to properly take an Abelian projection onto the $w_{\mu}^{xy}$ sector of Eq. \ref{eq:gcsDirac} to recover the gravitational Chern-Simons term with proper normalization as written in Eq. \ref{CS:Nematic}. 

\section{Conclusion}\label{section:conclusion}

In this paper, we discussed the proximate disordered phases and the continuous melting transitions driven by condensation of topological defects of crystals and liquid crystals coupled to a quantum Hall state. By developing the dual descriptions for the spatial order, the topological defects are shown to carry anyonic statistical angles and attached quantum numbers due to the Chern-Simons terms of the geometric deformations. Because the Chern-Simons terms attach the fluxes to the defects, the melting transitions are equivalent to a transition between different quantum Hall states. Thus the topological order as well as the symmetry may change at a single continuous transition. We also carefully examine the quantum numbers of the lattice defects and identify the proximate disordered phase of the liquid crystals. In the case of the crystal, we showed that the dislocation carries crystal momentum and thus the disordered phase of the crystal is again a crystal with the broken translational symmetry. We also studied the nematic phase and found that in cases where the electrons are strongly coupled to the nematic, the disclination carries fractional electric charge and an anyonic statistical angle. When the disclinations condense, the isotropic phase supports a {\it deconfined} bosonic excitation with fractional electric charge. Thus the isotropic phase and the nematic phase have different electromagnetic Hall effects.   

\acknowledgements
Authors thank A. Beekman, M. Stone, Z. Nussinov and E. Fradkin for helpful discussions. The authors acknowledge support from the NSF grant number DMR-1064319 (G.Y.C), NSF CAREER  DMR-1351895 (TLH), and U.S. DOE contract DE-FG02-13ER42001. 

\bibliography{Melting}
\end{document}